# Thermal Conductivity of 1,2-Ethanediol and 1,2-Propanediol Binary Aqueous Solutions at Temperature from 253 K to 373 K

Changyu Deng [1,2], Ke Zhang *,[1]

[1] School of Energy and Power Engineering, Xi'an Jiaotong University, Xi'an 710049, China

[2] Department of Mechanical Engineering, University of Michigan, Ann Arbor, MI 48109, United States

* Corresponding author, email: k.zhang@mail.xjtu.edu.cn

## Abstract

1,2-Ethanediol, 1,2-propanediol and their aqueous solutions are widely used as heat transfer fluids. Their thermal conductivity is a vital physical property, yet there are only few reports in literature. In this paper, thermal conductivity of binary aqueous solutions of the two glycols was measured using the transient hot wire method at temperature from 253.15 K to 373.15 K at atmospheric pressure. Measurement was made for six compositions over the entire concentration range from 0 to 1 mole fraction of glycol, namely, 0.0, 0.2, 0.4, 0.6, 0.8, and 1.0 mole fraction of glycol. The uncertainties of temperature and concentration measurement were estimated to be 0.01 K and 0.1 %, respectively. The combined expanded uncertainty of thermal conductivity with a level of confidence of 0.95 ($k = 2$) was 2 %. The second-order Scheffé polynomial was used to correlate the temperature and composition dependence of the experimental thermal conductivity, which was found to be in good agreement with the experiment data from the present work and other reports.

## 1 Introduction

1,2-Ethanediol and 1,2-propanediol are important chemicals in industry and scientific research. 1,2-Ethanediol, also known as ethylene glycol, is a commercial raw material for the manufacture of polyester fibers, chiefly polyethylene terephthalate, and can also be used as a humectant, plasticizer, softener, etc. 1,2-Ehanediol lowers the freezing point of water, therefore aqueous solutions of it are commercially applied as antifreezes. They are widely employed, for example, in motor vehicles, solar energy units, heat pumps, water heating systems, and industrial cooling systems [1]. 1,2-Propanediol, also called propylene glycol, is widely used in the manufacture of unsaturated polyester resins. It is a precursor of many polyether polyols used in urethane foam, elastomer, adhesives, and sealants industry. Its aqueous solutions are utilized in aircraft de-icing and anti-icing fluids because of its properties: low toxicity, ready biodegradability, and environmental acceptance [2]. Its solutions play an important role as heat transfer fluids and coolant agents owing to their ability to efficiently lower the freezing point of water and their low volatility. Although ethylene glycol solutions have better thermophysical properties than propylene glycol solutions, especially at lower temperature, the less toxic propylene glycol is preferred for applications involving possible human contact or where mandated by regulations [3].

The physical properties have to be known in process engineering and in heat exchanger design [4]. For instance, Najjar et al. demonstrated the influence of improved physical property data on calculated heat transfer rates and showed that the resulting error in heat transfer coefficients will be about 110 % if each of the estimated physical properties is 50 % higher [5, 6]. Among those thermal properties, thermal conductivity is essential to designing heat transfer and thermal energy storage systems, yet the thermophysical properties of these aqueous solutions are still scarce, especially at low temperature [3]. With regard to thermal conductivity at atmosphere, researchers were prone to investigate the two glycols together, probably due to their similar properties and applications. Literature on thermal conductivity of aqueous solutions of 1,2-ethanediol and 1,2-propanediol is summarized in Table 1 [7]. The data is sparse, especially considering the diversity of temperature and mass fraction. Therefore, more measurement is needed to meet the demands of industry and research.

**Table 1** Summary of literature on thermal conductivity of aqueous solutions of 1,2-ethanediol and 1,2-propanediol at atmospheric pressure[a]

| First author | Year | Temperature/K | Mass fraction/% |
|---|---|---|---|
| Bates, O. K. | 1945[8] | 293-383 | 10, 20, 30 ,40, 50, 60, 70, 80, 90 |
| Riedel, L. | 1951[9] | 233-373 | Same as above |
| Rastorguev, Yu. L. | 1966[10][b] | 313 | 20, 40, 60, 80 |
| Rastorguev, Yu. L. | 1967[11] | 313 | 78 |
| Vanderkool, W. N. | 1967[12] | 273-381[b]<br>273, 323, 373[c] | 20, 40, 60, 80 |
| Ganiev, Yu. A. | 1968[13][b] | 313 | 81 |
| Usmanov, I. U. | 1977[14][b] | 313 | 20, 40, 60, 80 |
| Bogacheva, I. S. | 1980[15][b] | 298-363 | 25, 50, 75 |
| Bohne, D. | 1984[6][b] | 280-470 | 25, 55, 75 |
| Grigrev, A. | 1985[16][b] | 302-454 | 14, 24, 62, 78 |
| Assael, M. J. | 1989[17] | 296-355 | 25, 50, 75 |
| Sun, T. | 2003[18][b],2004[19][c] | 299-442 | 25, 50, 75 |

a) Authors measured both glycols unless otherwise stated
b) 1,2-Ethanediol only
c) 1,2-Propanediol only

In this paper, thermal conductivity of binary aqueous solutions of 1,2-ethanediol and 1,2-propanediol was measured using the transient hot wire method at temperature from 253.15 K to 373.15 K covering the whole composition range at atmospheric pressure. The second-order Scheffé polynomial was used to correlate the temperature and composition dependence of the experimental thermal conductivity.

## 2 Experimental

## 2.1 Chemicals

The chemical samples of 1,2-ethanediol and 1,2-propanediol used in this work were analytical grade. Both of them had mass fraction purity of 99.0 % and were provided by Sinopharm Chemical Reagent CO, Ltd., China. Complete specification of chemical samples is listed in Table 2. Toluene was used to test our apparatus, as described later. Deionized and redistilled water was used throughout all of the experiments. All sample materials were used without further purification. In the experiments, the aqueous solutions were prepared by weighing, and then injected into the pressure vessel. An analytical balance (Mettler Toledo XS205) with an accuracy of ±0.1 mg was used to weigh the samples.

**Table 2** Specification of chemical samples

| Chemical | CAS number | Source | Initial mass fraction purity | Purification method |
|---|---|---|---|---|
| 1,2-Ethanediol | 107-21-1 | Sinopharm Chemical Reagent CO, Ltd., China | 99.0 | None |
| 1,2-Propanediol | 57-55-6 | Sinopharm Chemical Reagent CO, Ltd., China | 99.0 | None |
| Toluene | 108-88-3 | Tianjin Fuyu Industry of Fine Chemicals Co., Ltd., China | 99.5 | None |

## 2.2 Apparatus

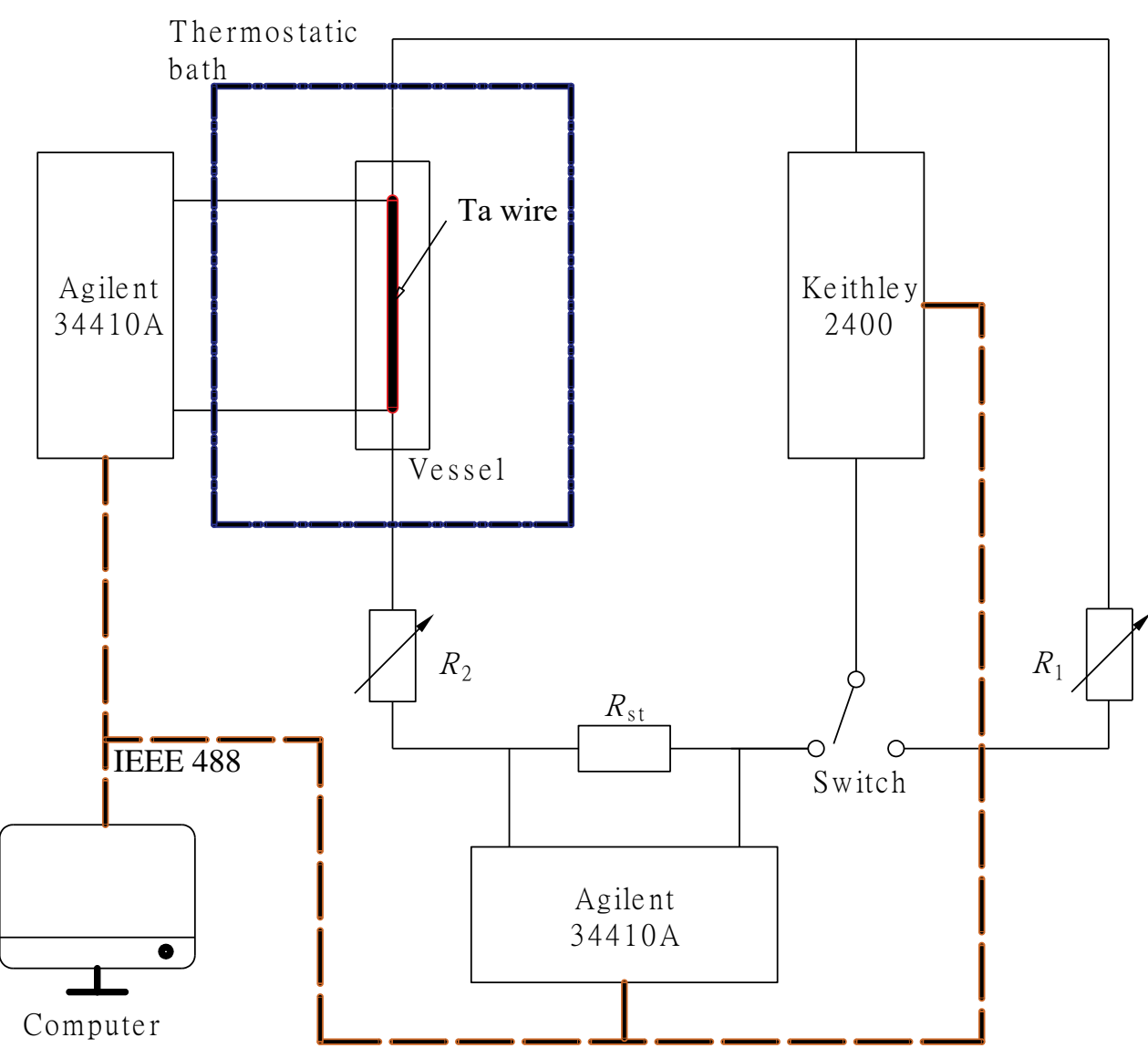

**Figure 1** Schematic diagram of transient hot-wire apparatus

The measurement of thermal conductivity was conducted by the transient hot-wire apparatus. The physical basis details of the transient hot-wire technique has been described elsewhere. The main structure of the apparatus, procedure of measurement, calibration procedure, and uncertainty assessment have been given in an earlier publication [20]. For experiments in present work, only the size of the pressure vessel and the multimeters of the data acquisition system were modified. Thus, a simple description is presented here.

The hot wire was a single tantalum wire with a diameter of 25 μm and a length of about 30 mm. The tantalum wire was anodized to form a layer of insulating tantalum pentoxide on its surface. The tantalum wire was enclosed by a stainless-steel vessel with a volume of about 20 mL. A schematic diagram of the transient hot-wire apparatus is presented in Figure 1. The power of the circuit was supplied by a Keithley 2400 sourcemeter. The resistance of the tantalum wire was obtained by measuring the current and voltage using two Agilent 34410 digital multimeters. All the data acquisition and instrument control were performed by a computer via the IEEE-488 interface.

The transient hot-wire apparatus was completely immersed in a thermostatic bath (Fluke, model 7037), whose temperature was measured with a platinum resistance thermometer. The temperature stability and uniformity of the bath were better than 10 mK.

The factors contributing to the uncertainty of thermal conductivity include temperature, pressure, mole fraction, the length of the tantalum wire, heating power, non-linearity of the $\Delta T$-ln$t$ curve, measurement repeatability and other negligible sources [21,22]. Measurement was performed 10 times at each temperature point. The relative standard deviation of the thermal conductivity was 0.9%, which represented the Type A uncertainty component. As for Type B, the uncertainty of temperature was 10 mK. The estimated fluctuation of the atmospheric pressure was ± 1 kPa. The relative standard uncertainty of mole fraction was estimated as 0.1%. The length of the tantalum wire was measured by a vernier caliper with an uncertainty of 0.02 mm. Considering the tension of the tantalum wire, the maximum uncertainty of the length measurement was estimated to be 0.1 mm, contributing to 0.19 % relative standard uncertainty. The heating power on the hot wire was calculated using two voltages (one for the hot wire, and the other for a standard resistor) measured by a 6 ½-digit resolution multimeter and the resistance of a standard resistor with an uncertainty of 0.002%. The non-linearity of the $\Delta T$-ln$t$ curve was reflected in the Type A uncertainty [21]. Therefore, all of the factors attributed to Type B resulted in an uncertainty of 0.3% in thermal conductivity measurements. Considering the aforementioned factors, the combined expanded uncertainty of the thermal conductivity with a level of confidence 0.95 ($k$=2) was 2 %.

The performance of the apparatus was tested by measuring the thermal conductivity of saturated liquid toluene from 273K to 373 K. Agreement with recommended values calculated by REFPROP software was within a maximum deviation of 1.20 % and an average absolute deviation of 0.61 %.

# 3 Results and Discussions

## 3.1 Pure liquids

Thermal conductivity of pure liquid 1,2-ethanediol and 1,2-propanediol is presented in Table 3.

**Table 3** Thermal conductivity of 1,2-ethanediol and 1,2-propanediol at pressure $p$=97 kPa.[a]

| 1,2-Ethanediol | | 1,2-Propanediol | |
|---|---|---|---|
| $T$ / K | $\lambda$ / W·m$^{-1}$·K$^{-1}$ | $T$ / K | $\lambda$ / W·m$^{-1}$·K$^{-1}$ |
| 263.23 | 0.2485 | 253.31 | 0.1979 |
| 273.20 | 0.2495 | 263.27 | 0.1974 |
| 283.09 | 0.2503 | 273.26 | 0.1969 |
| 293.06 | 0.2513 | 282.93 | 0.1964 |
| 312.91 | 0.2535 | 293.20 | 0.1962 |
| 332.63 | 0.2549 | 313.35 | 0.1955 |
| 352.54 | 0.2562 | 333.46 | 0.1951 |
| 363.09 | 0.2565 | 353.51 | 0.1945 |
| 372.48 | 0.2568 | 363.59 | 0.1943 |
| | | 373.59 | 0.1938 |

a) Standard uncertainty $u(p)$ =1 kPa, $u(T)$ = 10 mK; combined expanded uncertainty $U_c(\lambda)$ = 0.02·$\lambda$, with a coverage factor $k$ = 2.

For engineering application and further research, a continuous function of thermal conductivity is required.

Thermal conductivity of pure liquids was correlated as a function of temperature:[23]

$$\lambda_i = a_i \cdot T^2 + b_i \cdot T + c_i \,, \tag{1}$$

where $T$ denotes the absolute temperature of solutions in K, $a_i$, $b_i$ and $c_i$ are coefficients.

Data were fitted via the least-square method and correlation coefficients were obtained, shown in Table 4. The average absolute deviations (AAD) and the maximum absolute deviations (MAD) of the calculated thermal conductivity from experimental data are respectively 0.06 %, 0.09 % for 1,2-ethanediol, and 0.05 %, 0.10 % for 1,2-propanediol. The calculated values are in satisfying agreement with the experiment data.

**Table 4** Fitting coefficients for pure glycols

| Liquid | $a_i$ | $b_i$ | $c_i$ | MAD | AAD |
|---|---|---|---|---|---|
| 1,2-Ethanediol | $-3.7625\times10^{-7}$ | $3.1811\times10^{-4}$ | $1.9064\times10^{-1}$ | 0.09 % | 0.06 % |
| 1,2-Propanediol | $9.6913\times10^{-8}$ | $-9.2337\times10^{-5}$ | $2.1496\times10^{-1}$ | 0.10 % | 0.05 % |

Values of thermal conductivity of 1,2-ethanediol measured in this work are compared with reports by other researchers in Figure 2, and 1,2-propanediol in Figure 3. It can be seen that most data are within ±2 % of calculated lines. The maximum absolute deviations of 1,2-ethanediol and 1,2-propanediol from calculated values are 3.24 % and 2.55 % respectively.

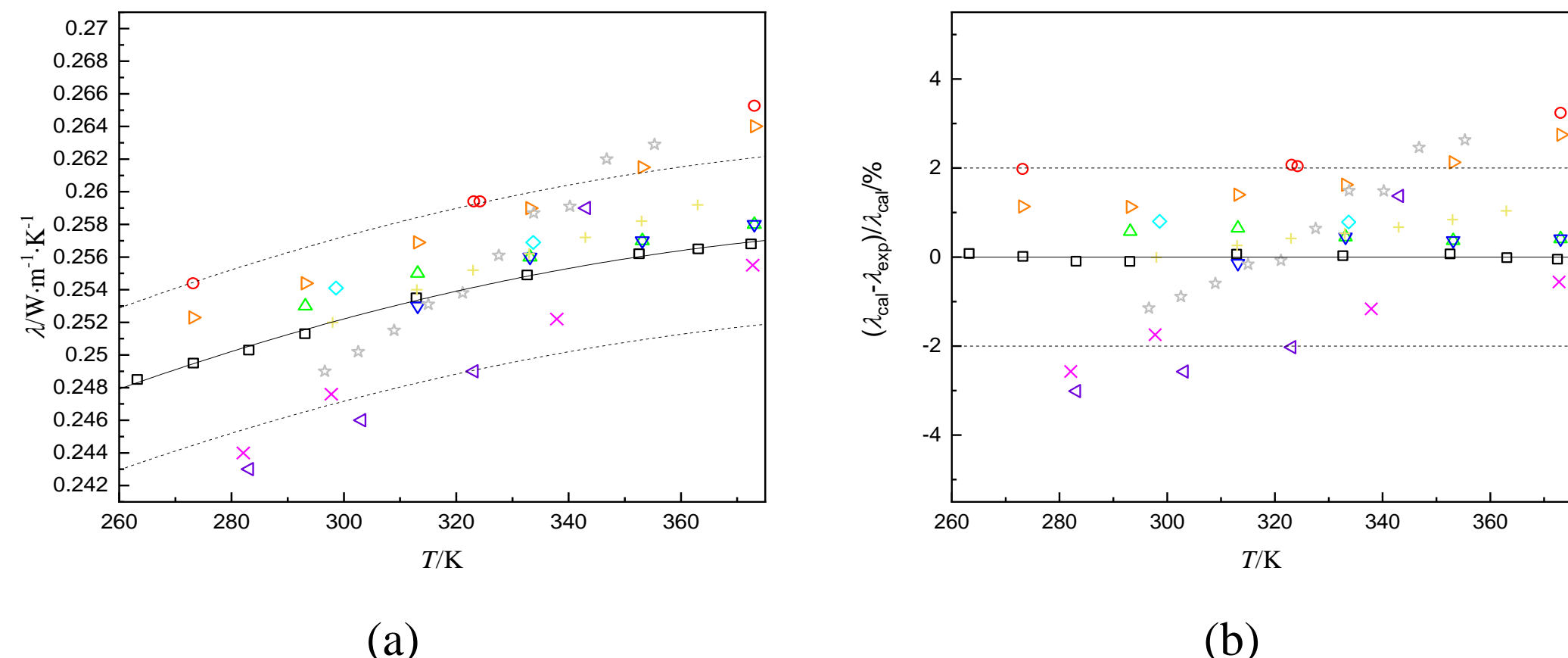

(a) (b)

**Figure 2** Comparison of thermal conductivity of 1,2-ethanediol. ○, Ref. [12]. △, Ref. [24]. ▯, Ref. [25]. ◇, Ref. [26]. ◁,Ref. [27]. ▷, Ref. [9]. +, Ref. [15] ×, Ref. [6]. ☆, Ref.[17]. □, present work. Solid line, calculated from correlation. Dash line, ±2 %.

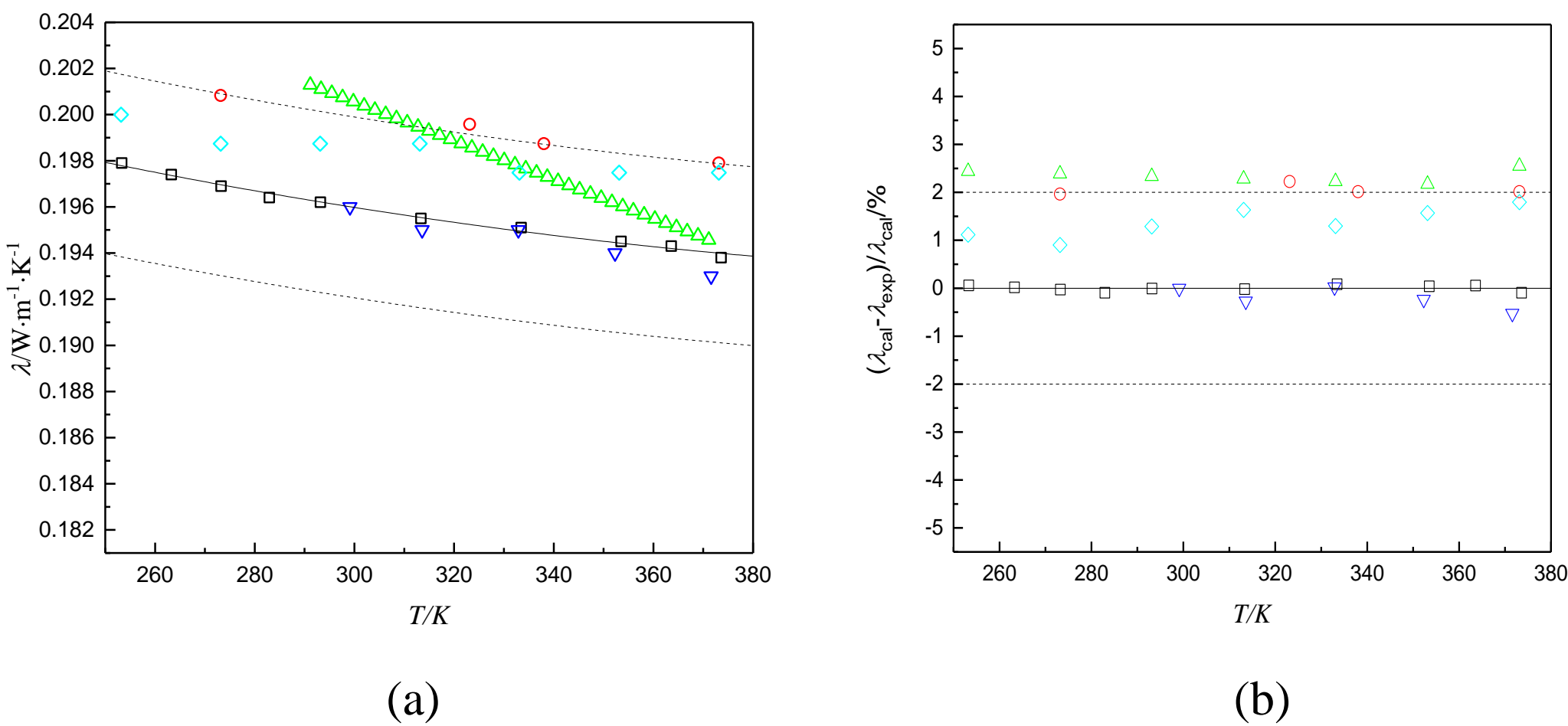

(a) (b)

**Figure 3** Comparison of thermal conductivity of 1,2-propanediol. ○, Ref. [12]. △, Ref. [28]. ▯, Ref. [19]. ▯, Ref.[9]. □, present work. Solid line, calculated from correlation. Dash line, ±2 %.

## 3.2 Mixtures

Experiment results of thermal conductivity with different temperature and fractions are presented in Table 5 (1,2-ethanediol) and Table 6 (1,2-propanediol).

**Table 5** Thermal conductivity of 1,2-ethanediol binary solutions at pressure $p$=97 kPa.[a] $x_1$ denotes mole fraction of 1,2-ethanediol; $w_1$ denotes mass fraction of 1,2-ethanediol.

| $T$/K | $\lambda$/W·m$^{-1}$·K$^{-1}$ | $T$/K | $\lambda$/W·m$^{-1}$·K$^{-1}$ |
|---|---|---|---|
| $x_1$=0.2008, $w_1$=0.4641 | | $x_1$=0.5980, $w_1$=0.8368 | |
| 253.27 | 0.3727 | 253.34 | 0.2802 |
| 263.09 | 0.3802 | 263.2 | 0.2820 |
| 272.91 | 0.3877 | 273.19 | 0.2848 |
| 282.70 | 0.3953 | 283.04 | 0.2874 |
| 292.68 | 0.4040 | 292.89 | 0.2897 |
| 312.40 | 0.4182 | 312.73 | 0.2951 |
| 332.22 | 0.4305 | 332.48 | 0.2993 |
| 352.05 | 0.4387 | 352.36 | 0.3028 |
| 362.06 | 0.4422 | 362.28 | 0.3040 |
| 372.03 | 0.4448 | 372.24 | 0.3048 |
| $x_1$=0.4014, $w_1$=0.6980 | | $x_1$=0.7992, $w_1$=0.9321 | |
| 253.63 | 0.3140 | 253.57 | 0.2542 |
| 263.33 | 0.3169 | 263.46 | 0.2553 |
| 273.19 | 0.3210 | 273.41 | 0.2564 |
| 283.12 | 0.3259 | 283.18 | 0.2572 |
| 292.94 | 0.3299 | 293.05 | 0.2585 |
| 312.75 | 0.3387 | 312.89 | 0.2612 |
| 332.49 | 0.3457 | 332.61 | 0.2633 |
| 352.41 | 0.3512 | 352.93 | 0.2654 |
| 362.35 | 0.3534 | 362.98 | 0.2660 |
| 372.39 | 0.3551 | 372.98 | 0.2664 |

a) Standard uncertainty $u(p)$ =1 kPa, $u(T)$ = 10 mK; relative standard uncertainty $u_r(x_1)$=0.001, $u_r(w_1)$=0.001; combined expanded uncertainty $U_c(\lambda) = 0.02\cdot\lambda$, with a coverage factor $k$ = 2.

Owing to the lack of fully developed thermal conductivity predictive models for liquid mixtures, empirical and semi-empirical correlation equations were considered in the literature. The second-order Scheffé polynomial is applied in this paper. With its simple forms, the polynomial is able to correlate binary data satisfactorily [29],

$$\lambda_m = \lambda_1 w_1^2 + \lambda_2 w_2^2 + 2\beta_{12} w_1 w_2, \tag{2}$$

with $\lambda_1$ and $\lambda_2$ thermal conductivity of pure liquids predicted by Eq.(1), and $\beta_{12}$ expressed by

$$\beta_{12} = \mathbf{A}_{12} + \mathbf{B}_{12} T \,. \tag{3}$$

Combining Eqs.(1)(2)(3), the thermal conductivity of binary solutions can be calculated by fractions of components and temperature. Thermal conductivity of pure water was obtained by IAPWS formulation[30]. The coefficients in these equations are presented in Table 7.

Table 6 Thermal conductivity of 1,2-propanediol binary solutions at pressure $p$=97 kPa.[a] $x_1$ denotes mole fraction of 1,2-propanediol; $w_1$ denotes mass fraction of 1,2-propanediol.

| $T$/K | $\lambda$/W·m$^{-1}$·K$^{-1}$ | $T$/K | $\lambda$/W·m$^{-1}$·K$^{-1}$ |
|---|---|---|---|
| $x_1$=0.2004, $w_1$=0.5143 | | $x_1$=0.6001, $w_1$=0.8638 | |
| 254.17 | 0.3358 | 253.30 | 0.2287 |
| 264.00 | 0.3403 | 263.16 | 0.2289 |
| 273.50 | 0.3466 | 273.09 | 0.2292 |
| 283.35 | 0.3529 | 283.02 | 0.2298 |
| 293.25 | 0.3588 | 292.99 | 0.2305 |
| 313.08 | 0.3691 | 312.64 | 0.2324 |
| 332.89 | 0.3792 | 332.78 | 0.2347 |
| 352.41 | 0.3849 | 352.58 | 0.2359 |
| 362.73 | 0.3878 | 362.73 | 0.2362 |
| 372.72 | 0.3890 | 372.56 | 0.2361 |
| $x_1$=0.4031, $w_1$=0.7405 | | $x_1$=0.8013, $w_1$=0.9446 | |
| 253.12 | 0.2643 | 253.01 | 0.2098 |
| 263.10 | 0.2652 | 263.23 | 0.2098 |
| 273.19 | 0.2670 | 274.26 | 0.2098 |
| 283.11 | 0.2687 | 284.30 | 0.2098 |
| 293.32 | 0.2720 | 294.28 | 0.2100 |
| 313.38 | 0.2773 | 313.85 | 0.2102 |
| 333.07 | 0.2817 | 333.77 | 0.2105 |
| 353.14 | 0.2849 | 353.26 | 0.2110 |
| 363.25 | 0.2866 | 363.14 | 0.2110 |
| 373.45 | 0.2869 | 373.20 | 0.2108 |

a) Standard uncertainty $u(p)$ =1 kPa, $u(T)$ = 10 mK; relative standard uncertainty $u_r(x_1)$=0.001, $u_r(w_1)$=0.001; combined expanded uncertainty $U_c(\lambda) = 0.02\cdot\lambda$, with a coverage factor $k$ = 2.

As the correlation error shown in Figure 4 and Figure 5, the average absolute deviations and the maximum absolute deviations of the calculated thermal conductivity of mixtures from experimental data are respectively 0.87 %,2.52 % for 1,2-ethanediol, and 0.53 %, 1.53 % for 1,2-propanediol. The calculated values are in satisfying agreement with the experiment data.

**Table 7** Fitting parameters for aqueous solutions. Subscript 1 denotes glycol, subscript 2 denotes water.

| Parameter | 1,2-Ethanediol | 1,2-Propanediol |
|---|---|---|
| $a_1$ | $-5.9997\times10^{-07}$ | $1.6981\times10^{-07}$ |
| $b_1$ | $4.5786\times10^{-04}$ | $-1.3731\times10^{-04}$ |
| $c_1$ | $1.6558\times10^{-01}$ | $2.1992\times10^{-01}$ |
| $a_2$ | $-9.2221\times10^{-06}$ | $-8.9967\times10^{-06}$ |
| $b_2$ | $7.1540\times10^{-03}$ | $7.0008\times10^{-03}$ |
| $c_2$ | $-7.0826\times10^{-01}$ | $-6.8217\times10^{-01}$ |
| $A_{12}$ | $2.3589\times10^{-01}$ | $2.3526\times10^{-01}$ |
| $B_{12}$ | $4.3606\times10^{-04}$ | $3.2318\times10^{-04}$ |
| MAD | 2.52 % | 1.53 % |
| AAD | 0.87 % | 0.53 % |
| Bias | -0.32 % | -0.21 % |

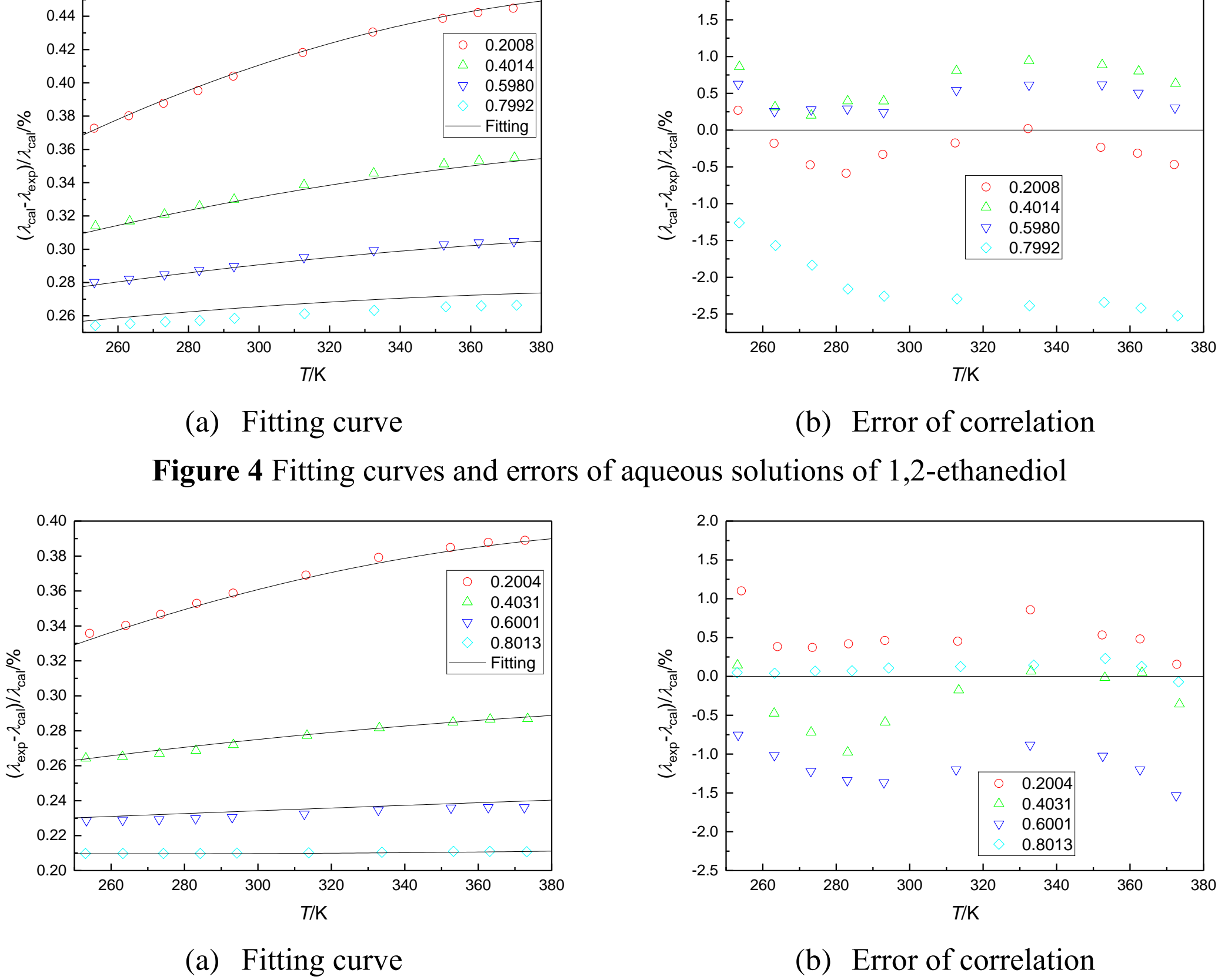

(a) Fitting curve (b) Error of correlation

**Figure 4** Fitting curves and errors of aqueous solutions of 1,2-ethanediol

(a) Fitting curve (b) Error of correlation

**Figure 5** Fitting curves and errors of aqueous solutions of 1,2-propanediol

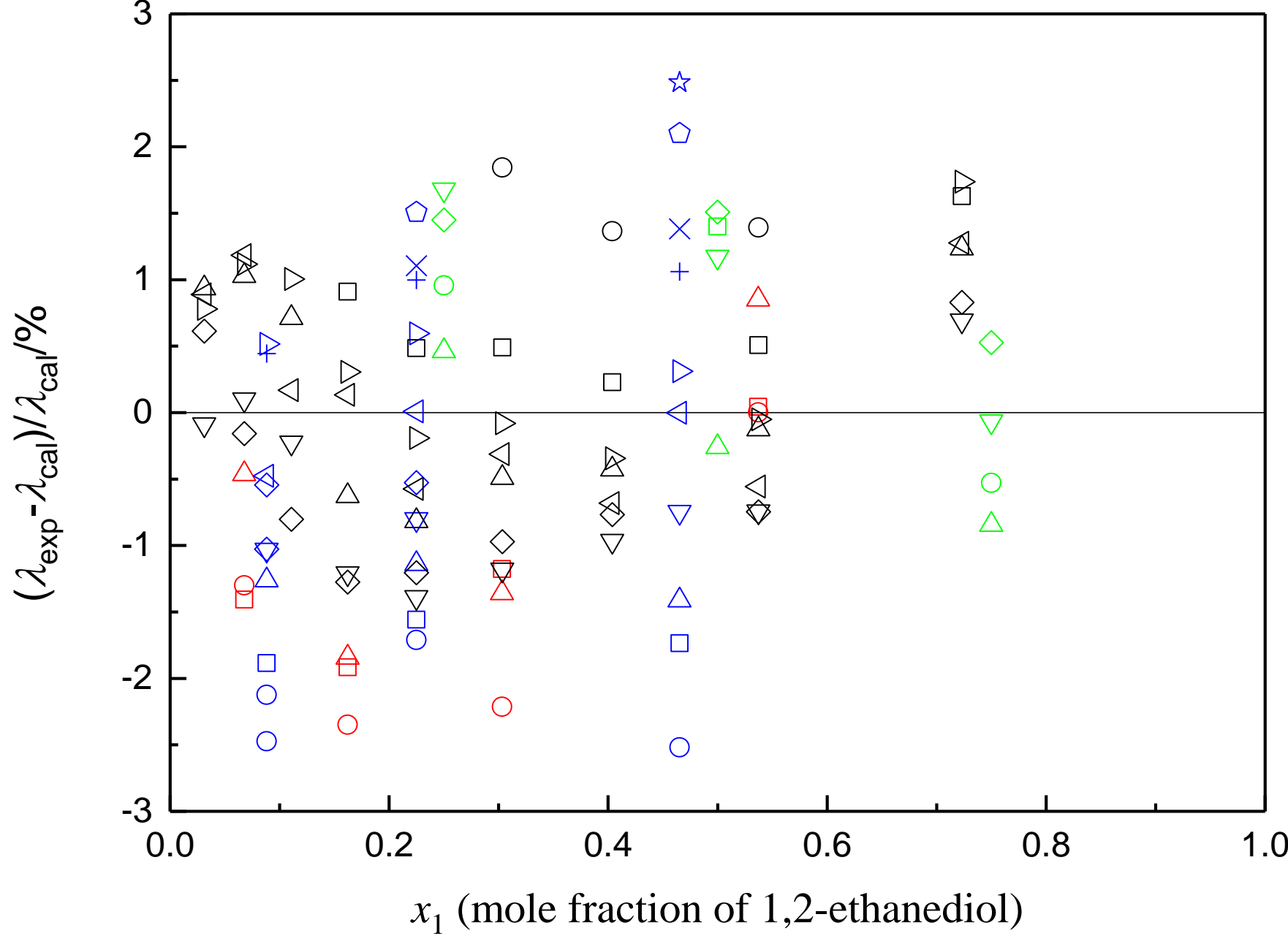

**Figure 6** Deviations of the thermal conductivity fitting equation of 1,2-ethanediol from experiment values of other investigators. Measurements at similar temperature (within ±2.5K) are put in the same temperature groups. Ref. [12]: ○273K, □323K, △373K. Ref. [17]: ○297K, □302K, △308K, ⍰314K, ◇319K, ◁325K, ▷329K, +335K, ×341K, ⬠347K, ☆355K. Ref [18]: ○301K, □312K, △324K, ▽348K, ◇372K. Ref[9]: ○233K, □253K, △273K, ▽293K, ◇333K, ◁353K, ▷373K.

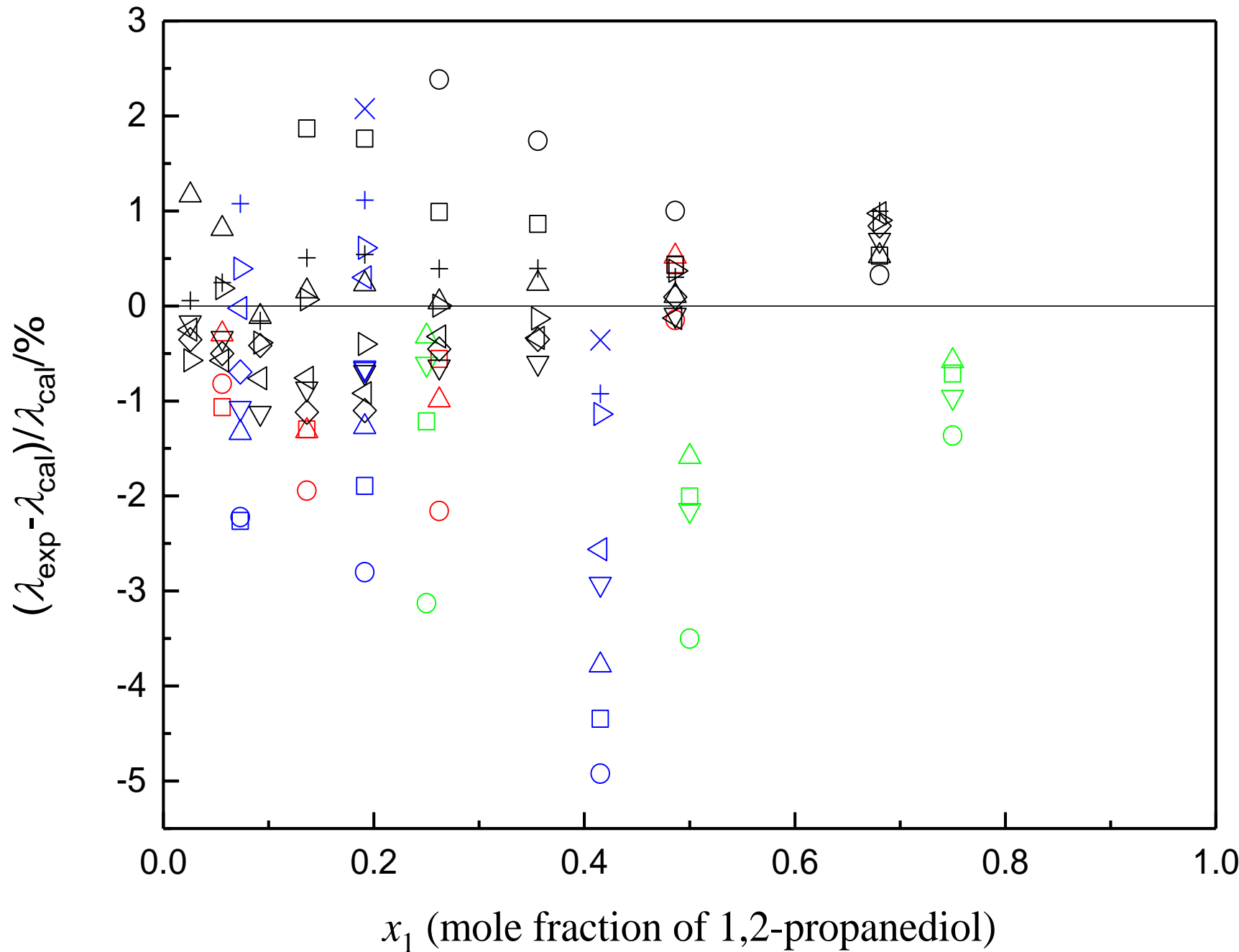

**Figure 7** Deviations of the thermal conductivity fitting equation of 1,2-propanediol from experiment values of other investigators. Measurements at similar temperature (within ±2.5K) are put in the same temperature groups. Ref. [12]: ○273K, □323K, △373K. Ref. [17]: ○297K, □302K, △310K, ⍰317K, ◇320K, ◁325K, ▷331K, +338K, ×345K. Ref.[19]: ○299K, □323K, △348K, ▽372K, ◇398K, ◁420K, ▷441K. Ref.[9]: ○233K, □253K, △273K, ▽293K, ◇313K, ◁333K, ▷353K, □373K.

The deviations of correlation from other authors' measurements are shown in Figure 6 and Figure 7. Most values are within ±3 % of the fitting equations, indicating the good agreement between the measurements of present work and others. The largest deviation is 2.52 % for 1,2-ethanediol, and 4.92 % for 1,2-propanediol.

## 4 Conclusions

Thermal conductivity of binary aqueous solutions of 1,2-ethanediol and 1,2-propanediol was measured using the transient hot wire method at temperature from 253.15 K to 373.15 K at atmospheric pressure, with mole fractions of glycol to be 0%, 20 %, 40 %, 60 %, 80 % and 100 % for both solutions. The combined expanded uncertainty of thermal conductivity was estimated to 2 % with a coverage factor of $k$=2. Thermal conductivity of pure liquids was correlated with temperature via second-order polynomial and was found to be in good agreement with other reports. The second-order Scheffé polynomial was used to correlate the temperature and composition dependence of the experimental thermal conductivity. The average absolute deviations and the maximum absolute deviations of those calculated values from the experimental data are 0.87 %, 2.52 % (1,2-ethanediol), and 0.53 %, 1.53 % (1,2-propanediol), respectively. Experiment values from other authors were compared with correlation functions to show good agreement.